\documentclass{WileyMSP-template}
\usepackage{textcomp, gensymb}
\usepackage{amssymb}
\usepackage{amsmath}
\usepackage{siunitx}
\usepackage{placeins}
\usepackage{algorithm}
\usepackage{algpseudocode}

\usepackage{mathtools}

\usepackage{caption}
\usepackage{subcaption}
\usepackage{indentfirst}
\usepackage{orcidlink}
\usepackage[misc]{ifsym}

\usepackage{parskip}
\usepackage{ragged2e}
\usepackage{hyperref}
\justifying

\newenvironment{varalgorithm}[1]
  {\algorithm[H]\renewcommand{\thealgorithm}{#1}}
  {\endalgorithm}

\begin{document}
\mathtoolsset{centercolon}

\pagestyle{fancy}

\title{A framework to compute resonances \\arising from multiple scattering}
\maketitle


\author{Jan~David~Fischbach\orcidlink{0009-0003-8765-8920} \Letter}
\author{Fridtjof~Betz}
\author{Nigar~Asadova}
\author{Pietro~Tassan\orcidlink{0009-0009-8493-6028}}
\author{Darius~Urbonas\orcidlink{0000-0001-9919-9221}}
\author{Thilo~Stöferle}
\author{Rainer~F.~Mahrt\orcidlink{0000-0002-9772-1490}}
\author{Sven~Burger\orcidlink{0000-0002-3140-5380}}
\author{Carsten~Rockstuhl\orcidlink{0000-0002-5868-0526}}
\author{Felix~Binkowski\orcidlink{0000-0002-4728-8887}}
\author{Thomas~Jebb~Sturges\orcidlink{0000-0003-1320-2843}}


\dedication{}

\begin{affiliations}
J. D. Fischbach, N. Asadova, Prof. Dr. C. Rockstuhl\\
Institute of Nanotechnology, Karlsruhe Institute of Technology\\
Email Address: fischbach@kit.edu
Dr. T. J. Sturges, Prof. Dr. C. Rockstuhl\\
Institute of Theoretical Solid State Physics, Karlsruhe Institute of Technology\\
F. Betz, Dr. S. Burger, Dr. F. Binkowski\\
Zuse Institute Berlin, Germany\\
P. Tassan, Dr. D. Urbonas, Dr. T. Stöferle, Dr. R. F. Mahrt\\
IBM Research Europe - Zurich, Switzerland\\
Dr. S. Burger\\
JCMwave GmbH - Berlin, Germany

\end{affiliations}



\begin{abstract}
Numerous natural and technological phenomena are governed by resonances. In nanophotonics, resonances often result from the interaction of several optical elements. Controlling these resonances is an excellent opportunity to provide light with properties on demand for applications ranging from sensing to quantum technologies. The inverse design of large, distributed resonators, however, is typically challenged by high computational costs when discretizing the entire system in space. Here, this limitation is overcome by harnessing prior knowledge about the individual scatterers that form the resonator and their interaction. In particular, a transition matrix multi-scattering framework is coupled with the state-of-the-art adaptive Antoulas–Anderson (AAA) algorithm to identify complex poles of the optical response function. A sample refinement strategy suitable for accurately locating a large number of poles is introduced. We tie the AAA algorithm into an automatic differentiation framework to efficiently differentiate multi-scattering resonance calculations. The resulting resonance solver allows for efficient gradient-based optimization, demonstrated here by the inverse design of an integrated exciton-polariton cavity. This contribution serves as an important step towards efficient resonance calculations in a variety of multi-scattering scenarios, such as inclusions in stratified media, periodic lattices, and scatterers with arbitrary shapes.

\end{abstract}



\section{Introduction}

Resonances play an integral role in many photonic applications. 
Deliberately tailored resonances can force the light in nanophotonic systems to obtain desired properties for specific applications. For example, they can promote and shape the emission of dipolar (quantum) emitters \cite{geQuasinormalModeApproach2014, renQuasinormalModesLocal2021, frankeQuantizationQuasinormalModes2019} and lasers \cite{stoferleUltracompactSiliconPolymer2010, renkBasicsLaserPhysics2017}. They are fundamental to various wavelength-selective devices like multiplexers and all-pass filters \cite{bogaertsSiliconMicroringResonators2012}.
The interplay among multiple resonances can lead to the emergence of complex behavior in observable quantities \cite{deckerHighEfficiencyDielectricHuygens2015}, including Fano resonances \cite{limonovFanoResonancesPhotonics2017} and the optical Vernier effect \cite{gomesOpticalVernierEffect2021}.

The interaction between coupled resonators can create a rich resonance spectrum beyond the original resonances of the uncoupled building blocks. By altering the shape and arrangement of subcells of such composite resonators, their spectrum can be tailored on demand. As a result, composite resonators play an enabling role in various fields of science and technology, including photonic crystal cavities, e.g., as interfaces for quantum computing \cite{frankeQuantizationQuasinormalModes2019, faraonCoherentGenerationNonclassical2008}, coupled-resonator optical waveguides \cite{yarivCoupledresonatorOpticalWaveguide1999} for broadband but sharp integrated optical filters \cite{melloniSynthesisDirectcoupledresonatorsBandpass2002}, high contrast grating cavities for silicon integrated lasers \cite{stoferleUltracompactSiliconPolymer2010}, and all-optical polariton transistors \cite{tassanIntegratedUltrafastAlloptical2024}. Similarly, resonant metamaterials formed by regular arrangements of subcells are of current interest \cite{pfeifferMetamaterialHuygensSurfaces2013, zhangChiralEmissionResonant2022, schulzRoadmapPhotonicMetasurfaces2024}. As such, they find applications as mirrors in, e.g., macroscopic Fabry-Perot resonators \cite{nymanDigitalTwinChiral2024}.

Electromagnetic resonances are known by many names across the literature, including quasi-normal modes, resonant states, natural modes, decaying states, and leaky modes. They are solutions to the time-harmonic source-free Maxwell equations obeying radiative boundary conditions. 
In passive systems with radiative and/or material losses, this eigenproblem 
is non-Hermitian. This results in complex-valued resonance frequencies $\tilde z$, with a negative imaginary part (in the sign convention at hand), 
which corresponds to the resonances' decay rate. The non-Hermitian character challenges the normalization of the modal fields that diverge away from the resonator. 
Various normalization strategies addressing this field divergence have been extensively debated. For the sake of brevity, we refer the interested reader to the following collection of recent articles \cite{lalanneLightInteractionPhotonic2018,kristensenModelingElectromagneticResonators2020,Binkowski2020prb,bothResonantStatesTheir2022, sauvanNormalizationOrthogonalityCompleteness2022, huangResonantLeakyModes2023,Betz2023pssa}.

Accurately and efficiently determining the resonances in complex photonic systems is an ongoing research challenge \cite{lalanneQuasinormalModeSolvers2019}. 
Proposed strategies include the harmonic inversion of time-domain simulations \cite{kristensenGeneralizedEffectiveMode2012, malhotraQuasinormalModeTheory2016}, perturbative methods \cite{dmitrievOpticalDownfoldingMethod2021} such as the so-called resonant state expansion \cite{muljarovResonantstateExpansionDispersive2016, muljarovResonantstateExpansionOpen2018}, and the use of Fredholm-type integral equations \cite{lassonThreedimensionalIntegralEquation2013, kristensenGeneralizedEffectiveMode2012}.
Nevertheless, most state-of-the-art methods for computing electromagnetic resonances follow one of two distinct approaches. The first approach directly treats the nonlinear eigenproblem (NEP) posed by the source-free Maxwell equations in dispersive media. This approach most commonly relies on some finite-element discretization paired with schemes to linearize the NEP \cite{demesyNonlinearEigenvalueProblems2020, yanRigorousModalAnalysis2018}.

In a second approach, the resonances can be found by considering a related inhomogeneous problem, where auxiliary sources are introduced into the system. This scattering problem will be formalized in Section~\ref{sec:scattering}. Crucially, its analytic continuation has singularities at complex frequencies that correspond directly to the resonances of the original system.
These singularities can be searched for via a variety of methods, e.g., gradient descent \cite{wiersigBoundaryElementMethod2002} and contour integration (with the Cauchy residue theorem) \cite{zollaFoundationsPhotonicCrystal2012,binkowskiPolesZerosNonHermitian2024}.




Owing to the fixed excitation frequency, this approach conveniently allows us to use any solver capable of evaluating the scattering problem at complex frequencies. Thus, it enables the use of further efficient problem-specific evaluation strategies, like the boundary element method \cite{wiersigBoundaryElementMethod2002, makitaloModesResonancesPlasmonic2014, powellInterferenceModesAllDielectric2017}, or the Fourier modal method 
\cite{tikhodeevQuasiguidedModesOptical2002, weissDerivationPlasmonicResonances2011, sauvanModalRepresentationSpatial2014}.
When treating the entire resonator as a single large system, one challenge pertinent to both approaches are the different length scales involved in many relevant compound resonators. Such treatment requires extensive memory and computing power. However, it is not necessary to consider the entire space occupied by a given photonic resonator when it is composed of multiple individual elements \cite{zollaFoundationsPhotonicCrystal2012, schwefelImprovedMethodCalculating2009, gagnonLorenzMieTheory2015}. 

\begin{figure}
     \centering
     \includegraphics[width=\linewidth]{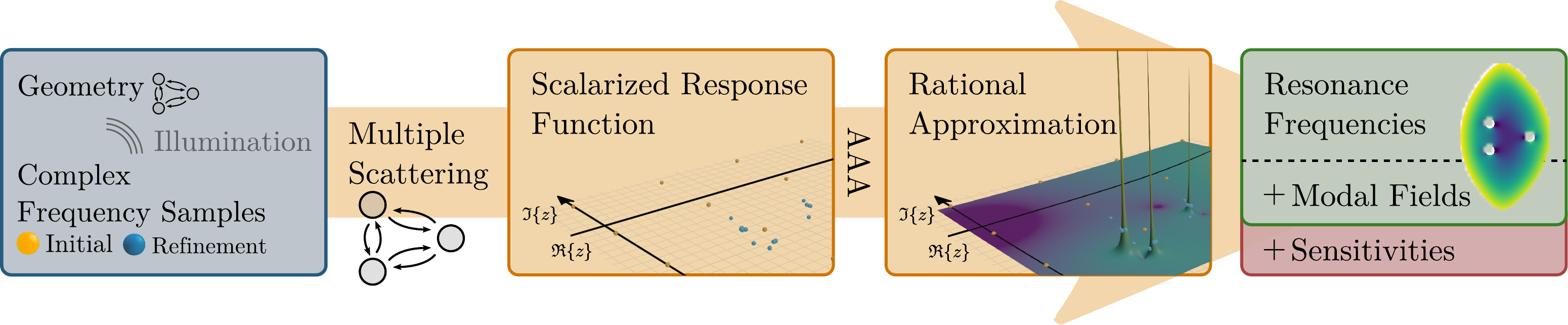}
     \caption{
     Schematic representation of the workflow of our proposed method: Single scatterers are arranged to form a resonator. The optical response is efficiently evaluated by multiple-scattering simulations at complex frequency samples $z_\mathrm{k}$. A scalarized response function $f(z_\mathrm{k})$ is computed from the simulated optical response. Then, the AAA algorithm is used to construct a rational approximation from the samples in the complex frequency plane. While the resonance frequencies can be directly obtained from the resulting approximation, a single additional scattering simulation is performed to obtain the corresponding modal fields. Moreover, the sensitivities of the resonance frequencies and their contribution to observable quantities (expressed in terms of its residue) w.r.t.~any degree of freedom characterizing the system can be obtained thanks to the automatic differentiation of the computational workflow. As a consequence, we can subject the resonances to the inverse design of the composite resonator (not shown).
     }
     \label{fig:schematic}
\end{figure}

Here, we instead address the subcells individually and \emph{a-posteriori} include their mutual coupling. Figure~\ref{fig:schematic} indicates schematically how we combine established methods for multiple-scattering resonance calculations. Specifically, we use analytical expressions for the electromagnetic wave propagation in a homogeneous background medium via the transition matrix formalism (TMF) \cite{zollaFoundationsPhotonicCrystal2012, beutelTreamsTmatrixbasedScattering2024}. To minimize the number of function evaluations while robustly recovering all relevant resonances, we choose the AAA algorithm \cite{nakatsukasaAAAAlgorithmRational2018} for pole finding \cite{betzEfficientRationalApproximation2024}. In contrast to contour integration methods, it allows us to freely add and remove frequency samples, thus supporting fine-grain adaptive refinement strategies, which is crucial to efficiently determining the rich resonance spectra of large ensembles of scatterers.

Additionally, we combine the established derivatives within the AAA algorithm \cite{betzEfficientRationalApproximation2024} with automatic differentiation \cite{griewankEvaluatingDerivativesPrinciples2008a} of the TMF to obtain gradients of the resonance frequencies.
Having these gradients at hand, we demonstrate the inverse design of resonances for a cavity of contemporary interest \cite{tassanIntegratedUltrafastAlloptical2024}. We demonstrate how to optimize the quality factor of a selected mode, achieving a level where it only remains limited by nonradiative losses. The considered system is an integrated cavity defined by dielectric posts in a stratified medium, arranged sequentially to form two mirrors. The stratified medium contains the ladder-type polymer MeLPPP that supports room-temperature exciton-polaritons \cite{plumhofRoomtemperatureBoseEinsteinCondensation2014}. These quasiparticles, which arise from the coupling of photons and excitons, form in the strong light-matter interaction regime when the polymer is placed in a wavelength-scale optical microcavity with a sufficiently high quality factor. These exciton-polariton devices have the potential to become crucial components of integrated all-optical computing platforms. To develop sophisticated polaritonic circuits and systems, it is essential to \emph{in-silico} optimize individual cavities and  arrangements of coupled cavities. 



\begin{figure}[ht]
    \centering    \includegraphics[width=0.9\linewidth]{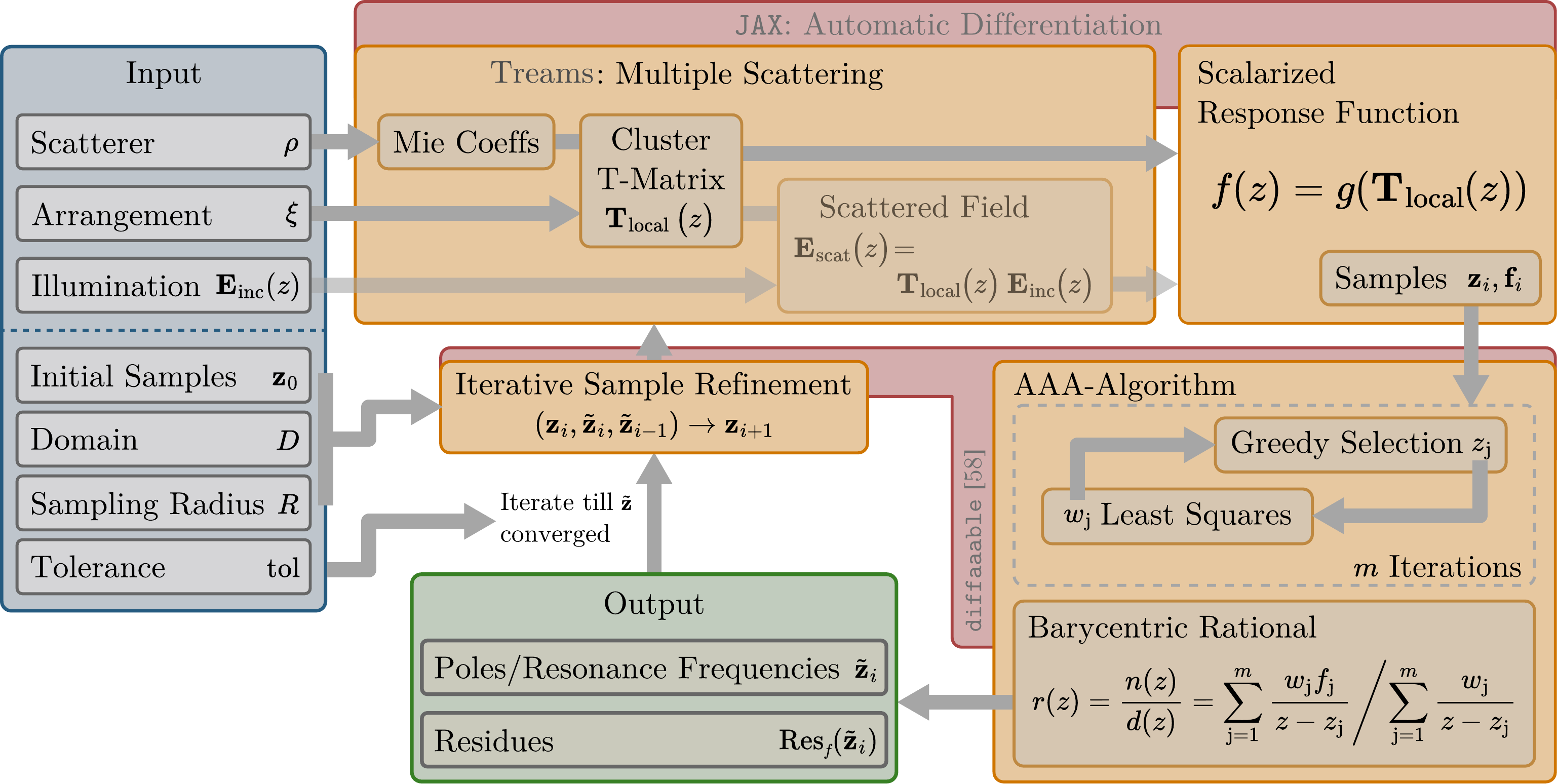}
    \caption{Flowchart detailing the procedure to determine resonance frequencies of a given arrangement $\xi$ of scatterers. The scattering problem is solved for an initial set of samples $\mathbf{z}_0$. 
    The solutions are scalarized as chosen by the user (via the choice of $g(\mathbf{T}_\mathrm{local})$). The sample pairs are aggregated into the vectors $\mathbf{z}_i$ and $\mathbf{f}_i$. The AAA algorithm is then used to estimate the pole positions $\tilde{\mathbf{z}}_0$, equivalent to the complex resonance frequencies. To ensure the accuracy of these resonance frequencies, samples are iteratively added in the vicinity of poles, which are selected according to Algorithm~\ref{algo:refinement}. User inputs are indicated in the blue box to the left. Methods to calculate sensitivities of the resonances (indicated in red) are further discussed in Section \ref{sec:inverse_design}.}
    \label{fig:detailed_schematic}
\end{figure}

\section{Method}
In the following, a detailed description of the proposed computational workflow will be provided. Figure~\ref{fig:detailed_schematic} guides the reader through our approach, starting with the central multiple-scattering formalism and continuing clockwise.

\subsection{The Scattering Problem: Transition Matrix Formalism} \label{sec:scattering}

Instead of directly treating the NEP, we will take the second approach mentioned in the introduction and identify the resonances as singularities of the time-harmonic Maxwell equations including sources \cite{binkowskiPolesZerosNonHermitian2024}:

\begin{equation}
\nabla\times\mu^{-1}(\mathbf{r}, z)\nabla\times\mathbf{E}(\mathbf{r}, z)-z^2\varepsilon(\mathbf{r}, z)\mathbf{E}(\mathbf{r}, z)=iz\mathbf{J}(\mathbf{r}, z)\, .\label{eq:Jscat}
\end{equation}

Here, $\mathbf{E}(\mathbf{r}, z)$ is the position dependent total electric field resulting from the excitation by the source current $\mathbf{J}(\mathbf{r}, z)$ at the complex frequency $z$, in and around the entire resonator. The $e^{-izt}$ sign convention is used in this contribution.
 


Let us consider a resonator consisting of an ensemble of individual optical elements (denoted as scatterers) embedded in a homogeneous background. With all sources located outside the individual scatterers, their individual response can be described in terms of the scattered field $\mathbf{E}_\mathrm{scat}$ resulting from some incident field $\mathbf{E}_\mathrm{inc}$ \cite[Annex~2]{lalanneLightInteractionPhotonic2018}.
The incident and scattered fields can be expanded into a set of basis functions that are elementary solutions to Maxwell's equations in the background medium. The vector cylindrical waves (VCW; 2D) and vector spherical waves (VSW; 3D) each form a complete basis of such solutions.
By expressing the incident and scattered field at a complex frequency $z$ as coefficient vectors in such a basis, the scattering response from an individual scatterer can be encapsulated into the operator $\mathbf{T}(z)$ defined by:

\begin{equation}
\mathbf{p}(z)=\mathbf{T}(z)\mathbf{a}(z)\, ,\label{eq:T}
\end{equation}

where $\mathbf{a}(z)$ are the incident field coefficients and $\mathbf{p}(z)$ are the scattered field coefficients. 
To make the expressions numerically tractable, the multipolar basis of $\mathbf{a}(z)$ and $\mathbf{p}(z)$ is typically truncated by a maximum multipolar order $m_\mathrm{max}$ (2D) or degree $l_\mathrm{max}$ (3D). Truncating the multipolar order permits the expression of $\mathbf{T}(z)$ as a finite matrix. The error introduced by the truncation is discussed in the SI Section~S3. 
Determining $\mathbf{T}(z)$ is at the core of solving the scattering problem for a single scatterer. A variety of methods to numerically evaluate $\mathbf{T}(z)$ for arbitrarily shaped scatterers exist \cite{demesyScatteringMatrixArbitrarily2018, fruhnertComputingTmatrixScattering2017, asadovaTmatrixRepresentationOptical2024}. Analytical expressions for $\mathbf{T}(z)$ are available for high-symmetry scatterers, such as infinitely extended cylinders (circles in 2D) and spheres \cite{bohrenAbsorptionScatteringLight1998}. These are commonly referred to as (generalized) Mie coefficients (first block of ``Multiple Scattering" in Figure~\ref{fig:detailed_schematic}). As shown in the SI Section~S1.1, the Mie coefficients can be evaluated at complex frequencies $z$ and are meromorphic with respect to $z$ in the relevant section of the complex plane. 

By considering the scattered field of one scatterer as part of the incident field to all other scatterers arranged to form the resonator, and vice versa, one can fully describe the effect of multiple scattering in a cluster (central block of ``Multiple Scattering" in Figure~\ref{fig:detailed_schematic}) \cite{beutelTreamsTmatrixbasedScattering2024}:
\begin{equation}
\mathbf{p}_\mathrm{local}(z)=\underbrace{\left[1-\mathbf{T_\mathrm{diag}}(z)\mathbf{C}^{(3)}(z)\right]^{-1}\mathbf{T_\mathrm{diag}}(z)}_{\mathbf{T_\mathrm{local}}(z)}\mathbf{a}_\mathrm{local}(z)\, .
\label{eq:multiscat}
\end{equation}

Here, $\mathbf{T}_\mathrm{diag}(z)$ is the block diagonal matrix constructed from the $\mathbf{T}(z)$ for each isolated constituent scatterer. $\mathbf{C^{(3)}}(z)$ encapsulates the translation coefficients to transform the scattered field from one scatterer into the basis of the other scatterers. We note that $\mathbf{C^{(3)}}(z)$ is holomorphic in $z$ in the relevant domain (SI~1.2). 
The field coefficients and the total transition matrix in Equation~\ref{eq:multiscat} are expressed in the joint local basis of the constituent particles, containing VSWs/VCWs around all scatterers. A basis change to a global basis with VSWs/VCWs around a single shared origin is possible.

The collective response of the cluster to every possible excitation is encapsulated in the resulting $\mathbf{T}_\mathrm{local}$-matrix. As a consequence, $\mathbf{T}_\mathrm{local}(z)$ contains all resonances that interact with the environment as poles of its entries. The fact that the transition matrix, and equivalently the scattering matrix $\mathbf{S} = \mathbf{1} + 2\mathbf{T}$ \cite{watermanSymmetryUnitarityGeometry1971}, contain the resonances of the object they describe is well established \cite{alpeggianiQuasinormalModeExpansionScattering2017, wuIntrinsicMultipolarContents2020, weissHowCalculatePole2018, benzaouiaQuasinormalModeTheory2021}. 
The $\mathbf{T}$-matrix permits various scalarization strategies to arrive at a scalar meromorphic response function $f(z)$, later used for pole finding. This choice determines the residues, which can help to emphasize poles of interest. Contrary to the scattering problem in terms of a source current $J(\mathbf{r}, z)$, within the $\mathbf{T}$-matrix formalism it is optional to select a source (indicated by the translucent third block of ``Multiple Scattering" in Figure~\ref{fig:detailed_schematic}, also see Section~\ref{sec:many_modes} for possible benefits). Alternatively, the transition matrix, which already contains the response to all possible excitations, can be scalarized by some linear function $g(\mathbf{T}_\mathrm{local}(z))$, for example, linear combinations of selected entries or the determinant of $\mathbf{T}_\mathrm{local}(z)$.


\subsection{Complex Pole Finding: AAA Rational Approximation}\label{sec:pole_finding}

With this strategy to efficiently evaluate the scalarized response function $f(z)$ in place, it remains to identify its poles. In principle, any of the aforementioned pole-finding methods could be applied. We use the recently published AAA algorithm for rational approximation \cite{nakatsukasaAAAAlgorithmRational2018, betzEfficientRationalApproximation2024} due to its multiple benefits:

\begin{enumerate}
    \item It is less dependent on \emph{initial values} to identify all relevant poles. Methods based on gradient descent particularly suffer from this limitation. 
    
    \item It provides excellent \emph{convergence} with the number of frequency samples.
    
    \item The convergence characteristics are complemented by the freedom to add and remove samples. As such, it permits \emph{refinement strategies} to iteratively add a small number of samples around relevant poles (see Algorithm~\ref{algo:refinement}). This feature distinguishes the approach from contour integral methods, as numerical quadratures require specific sampling strategies for a comparable convergence. 
\end{enumerate}

 

The barycentric rational form is at the heart of the AAA algorithm:
\begin{equation}
r(z)=\frac{n(z)}{d(z)}=\sum_{j=1}^m\left.\frac{w_jf_j}{z-z_j}\right/\sum_{j=1}^m\frac{w_j}{z-z_j} \approx f(z)\, .
\label{eq:baryrat}
\end{equation}
It possesses the interpolation property of reproducing $\lim_{z\to z_\mathrm{j}}r(z) = f_\mathrm{j}$.
Using this property, the approximation is iteratively constructed by greedily aggregating a selection of considered samples $z_j$ from the set of all samples (with elements $z_\mathrm{k}$). In every step, the corresponding weights $w_j$ are chosen to minimize the residual error at the remaining samples (schematically indicated in ``AAA-Algorithm" in Figure~\ref{fig:detailed_schematic}).
The poles and their residues are easily accessible from the constructed approximation (see ``Output" in Figure~\ref{fig:detailed_schematic}). However, they might not yet have the desired accuracy at this point.

The third property of the AAA algorithm mentioned above allows us to freely place additional samples near estimated poles to refine them \cite{betzEfficientRationalApproximation2024}.
Here, we generalize this idea to an iterative aggregation scheme for $z_\mathrm{k}$ (see Algorithm \ref{algo:refinement} indicated in Figure~\ref{fig:detailed_schematic} by ``Adaptive Refinement"). This modification allows us to efficiently and simultaneously find a \emph{large} number of relevant poles (see Section~\ref{sec:many_modes}). It avoids overly refining poles that have already been localized to a satisfactory degree.  

    
        
    

\begin{figure}[ht]
\centering
   \begin{minipage}{.38\linewidth}
    \begin{varalgorithm}{ISR}
    \caption{Iterative Sample Refinement}\label{algo:refinement} 
    \begin{algorithmic}
    \State $\mathbf{z}_\mathrm{add} \gets \mathbf{z}_{0}$
    \While{$\mathbf{z}_\mathrm{add}$ not empty} 
        \State $\mathbf{z}_{l} \gets \mathbf{z}_{l-1} \oplus \mathbf{z}_\mathrm{add}$
        \State $\mathbf{f}_{l} \gets \mathbf{f}_{l-1} \oplus [f\left(z) \, \mathbf{for} \, z \in \mathbf{z}_\mathrm{add}\right)]$
    
        \State $\tilde{\mathbf{z}}_{l} \gets \mathrm{AAA}(\mathbf{z}_{l}, \mathbf{f}_{l})$
        
        \State $\mathbf{z}_\mathrm{add} \gets [\tilde z + R e^{i\varphi_\mathrm{rand}} \, \mathbf{for} \, \tilde z \in \tilde{\mathbf{z}}_l \cap \mathcal{D}$ \\ \hspace{1.8cm} $\mathbf{if} \; \mathrm{min}|\tilde z - \tilde{\mathbf{z}}_{l-1}|>\mathrm{tol}]$
        
    
        \State $l \gets l+1$
    \EndWhile
    \end{algorithmic}
    \end{varalgorithm}
   \end{minipage}
   \caption*{Algorithm ISR: Algorithm to reduce the number of function evaluations needed for convergence of the poles estimated via the AAA approximation. Starting from a vector of initial samples $\mathbf{z}_0$ a first AAA approximation is constructed and the corresponding poles $\tilde{\mathbf{z}}$ are determined. In the first iteration, one additional sample per estimated pole is placed in the pole's vicinity (at a fixed distance $R$ and random direction $\varphi_\mathrm{rand}$ from the pole) if it lies within the domain of interest $\mathcal{D}$. In successive iterations, additional samples are only placed if the corresponding pole is not matched by any pole in the previous iteration (within some specified tolerance), i.e., the pole's position has not yet converged. The algorithm terminates after all poles have converged.
   $l$ is the iteration number. $\mathbf{z}_l$, $\mathbf{f}_l$, and $\tilde{\mathbf{z}}_l$ are vectors containing the frequency samples, corresponding function values, and estimated poles, respectively. $\mathbf{z}_{-1}$, $\mathbf{f}_{-1}$, and $\tilde{\mathbf{z}}_{-1}$ are empty. The $\oplus$ symbol indicates the concatenation of vectors.}
\end{figure}

As discussed in \cite{betzEfficientRationalApproximation2024}, the samples not incorporated in $r(z)$ can be used to find the \emph{derivatives} of $w_j$ w.r.t. the sampled values $\mathbf{f}$. A Python implementation of the derivatives compatible with JAX automatic differentiation is provided with this contribution \cite{Fischbach2024} (represented by the lower red box in Figure~\ref{fig:detailed_schematic}). In conjunction with the automatically differentiable TMF (represented by the upper red box in Figure~\ref{fig:detailed_schematic}), this allows us to compute the gradients of the poles (i.e., resonance frequencies) relative to all degrees of freedom in the design. This ability is key to performing gradient-based inverse design as described in Section~\ref{sec:inverse_design}. Implementation details for the AAA derivatives are laid out in Section~S2 of the SI. 








\section{Results}
Due to their strong and ultrafast nonlinearity, exciton-polaritons hold great potential as a platform for all-optical logic operations. However, chip integration is required to scale up and leverage their ultra-fast switching in complex circuits. Ongoing efforts use cavities formed by high contrast grating (HCG) meta-mirrors for this purpose \cite{tassanIntegratedUltrafastAlloptical2024}.
These cavities are fabricated in an SOI platform by nanostructuring the top-silicon layer, which is optically isolated from the silicon substrate by a buried oxide layer. The organic ladder-polymer MeLPPP is spin-coated onto the cavity and encapsulated by a layer of $\mathrm{Al_2O_3}$ deposited by atomic layer deposition (see Figure~\ref{fig:optimization_sensitivities}a). As the encapsulated polymer has a higher refractive index than the SiO$_2$ below and the air above, it supports vertically confined slab modes. Lateral photonic confinement is realized by the lithographically defined silicon nanopillars. 

The examples we will treat throughout the rest of this contribution are chosen to showcase selected aspects of the proposed method, while introducing relevant concepts, later used in the inverse design of such integrated polariton cavities. 
We start by examining the computational cost and convergence of finding the resonance frequencies for a small number of coupled scatterers. We then demonstrate how modal fields can be evaluated and go over methods for emphasizing or suppressing resonances using an appropriate choice of excitation. 
In particular, we study a macroscopic cavity formed by metamaterial mirrors, robustly recovering its rich spectrum and modal fields. Tracking relevant modes becomes particularly important when using the available sensitivities for gradient-based inverse design, which is discussed last.

\subsection{Resonance Frequencies}\label{sec:convergence_poles}
\begin{figure}[ht]
    \centering
    \includegraphics[width=\linewidth]{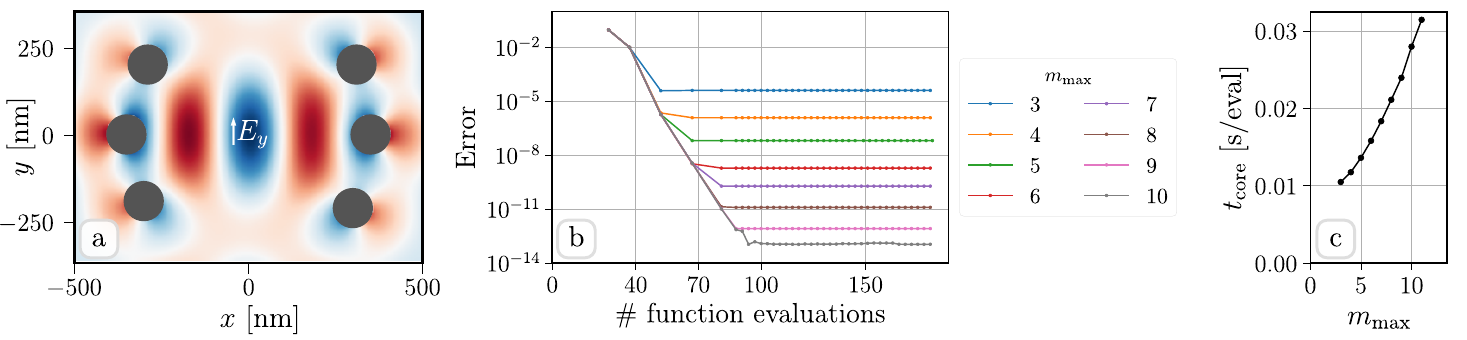}
    \caption{Convergence study of the proposed method for a small resonator made from an assembly of cylindrical scatterers. 13 resonances between \qty{1.6}{eV} and \qty{2.7}{eV} are considered simultaneously a) Dominant field component of the resonance with the lowest losses located at $\tilde{z} \approx \complexqty{2.142-0.074i}{eV}$. The field is not evaluated inside the cylindrical scatterers, where the expansion of the scattered field into radiating VCWs is not valid. b) Execution time on a single core per function evaluation depending on the truncation order $m_\mathrm{max}$ of the cylinders' $\mathbf{T}$-matrix. c) Convergence of the error in resonance frequencies with an increasing number of frequency samples, given as the maximum relative error (here denoted as ``Error") of the relevant poles compared to reference poles evaluated with $m_\mathrm{max}=11$. To clearly show the convergence behavior, the samples were selected by Algorithm \ref{algo:refinement} modified to yield at least three new samples (close to the most volatile poles) for 38 iterations.}
    \label{fig:convergence_poles}
\end{figure}

Let's consider the first example resonator made of 6 dispersive silicon pillars (the parameters of the Lorentz oscillator model are provided in SI~S4) embedded in a background permittivity $\varepsilon_{r,\mathrm{bg}} = \complexnum{2.9 + 0.001i}$ as shown in Figure~\ref{fig:convergence_poles}a. The pillars have a radius of $R=\qty{55}{nm}$ and are positioned such that the center-to-center distance of the central cylinders is \qty{700}{nm}. The upper and lower posts are displaced by \qty{200}{nm} in the $y$-direction and \qty{50}{nm} inwards along the $x$-direction. 
Light is scattered between the cylinders leading to the formation of collective resonances. Figure~\ref{fig:convergence_poles}a illustrates the electric field distribution of one such resonance with frequency $\tilde{z} \approx \complexqty{2.142-0.074i}{eV}$. In the following, we will analyze the accuracy and computational cost of determining multiple resonance frequencies simultaneously using our method.  


Figure~\ref{fig:convergence_poles}b shows how the error of the resonance frequencies converges while adaptively adding frequency samples according to Algorithm~\ref{algo:refinement}. The number of samples needed to obtain satisfactory accuracy has to be scaled by the computational cost to evaluate $f(z)$. 
The single-threaded time spent to evaluate the scattering problem at one complex frequency is displayed in Figure~\ref{fig:convergence_poles}c for different truncation orders $m_\mathrm{max}$. 
The shown error is the maximum relative error amongst 13 resonances falling in the considered domain of the complex frequency plane between \complexqty{1.6-0.4i}{eV} and \complexqty{2.7+0.1i}{eV}. 
The error is dominated by the $\mathbf{T}$-truncation from ca. 70 samples onwards for moderate $m_\mathrm{max}$. 

In Section~S3 of the SI, we compare the found resonances to the results of a state-of-the-art eigensolver based on Arnoldi iteration, which is part of the commercially available finite element method (FEM) software package JCMsuite. The resonance frequencies found by the AAA approximation converge to the reference solution with a remaining maximum relative error of $\sim 10^{-9}$, which is within the expected accuracy of the FEM solution (see Figure~S1b). %

Considering the combination of Figures~\ref{fig:convergence_poles}b and ~\ref{fig:convergence_poles}c reveals the trade-off between computational effort and accuracy of our method. For example, a desired accuracy of \num{1e-9} would require $m_\mathrm{max}\geq7$ with approximately \num{85} frequency samples, resulting in a total single-threaded execution time of less than \qty{2}{s}. 
Comparing that to the Arnoldi iteration, which takes more than \qty{200}{s} for the same accuracy (Figure~S1b), we observe a computational advantage of about two orders of magnitude for the system at hand. For larger systems, especially in 3D, our proposed method should offer even further savings in computational cost. Furthermore, the computational complexity of the TMF generally reduces for a wider spacing of the scatterers (because a lower $m_\mathrm{max}$ is needed to capture near-field interaction accurately). 

\subsection{Modal Fields}

Besides its resonance frequency, a mode is characterized by the corresponding field distribution. As shown in \cite{betzEfficientRationalApproximation2024}, it is possible to construct the modal fields by a weighted superposition of the fields at the sample frequencies $z_\mathrm{k}$. This is equally possible when the fields are expressed in the VSW/VCW-basis. However, in contrast to \cite{betzEfficientRationalApproximation2024}, we do not necessarily calculate $f(z)$ as a linear function of the scattered field.
Therefore, we evaluate the modal fields using a single scattering simulation close to the pole of interest. The response to any source that couples to a non-degenerate mode will take the form of such mode when the excitation frequency approaches the resonance frequency of the mode~\cite{baiEfficientIntuitiveMethod2013}. We further note that for increased efficiency, only the scattered field distribution has to be evaluated, as it will dominate over the incident field when approaching the pole.

As demonstrated numerically in Section~S3, the modal fields converge to the state-of-the-art reference eigensolver with the exception of the fields very close to the cylinders. The observed error has more than $m_{max}$ oscillations around the scatterers and decreases in magnitude when increasing $m_\mathrm{max}$. It is thus attributed to the error introduced by the truncation in the multipolar order. The fields shown in Figure~\ref{fig:convergence_poles}a and in further figures were calculated with that approach.

\subsection{Selective Excitation}
\label{sec:many_modes}

\begin{figure}
    \centering
    \includegraphics[width=\linewidth]{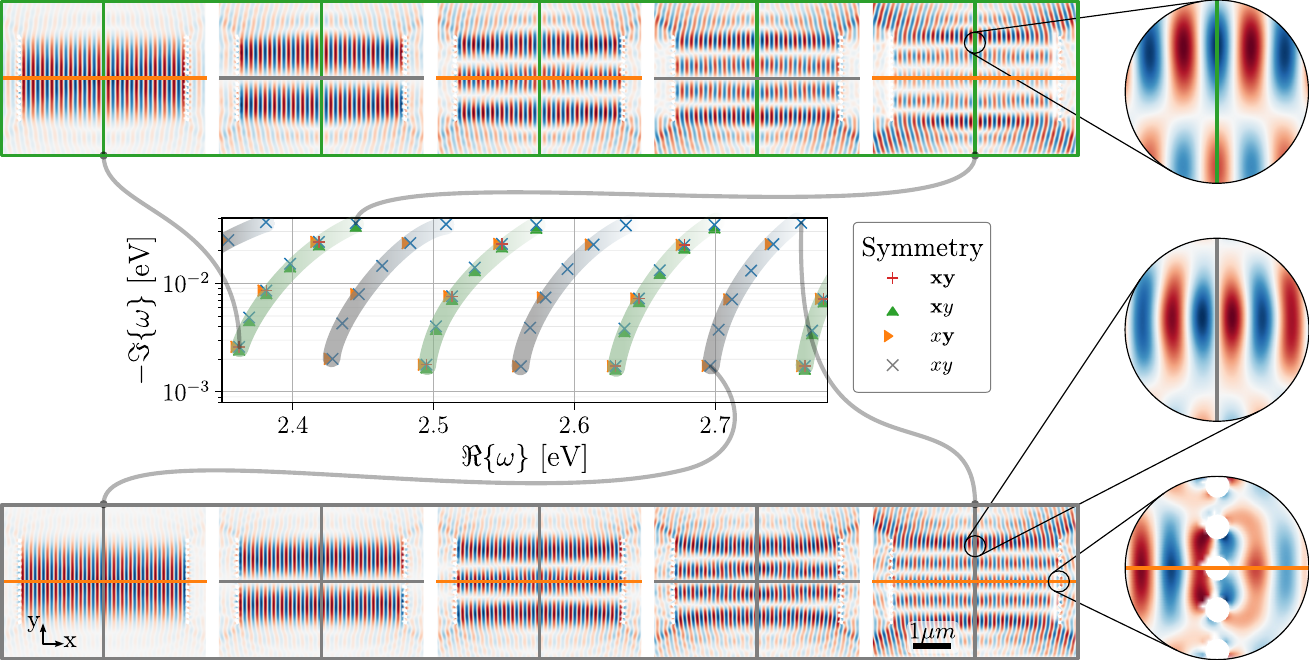}
    \caption{Modes of a large composite resonator made of 30 cylinders. The central figure shows the complex resonance frequencies of the modes, and the plots above and below show the corresponding field distribution of a few selected modes. Only the dominating $E_y$ component is shown in these field plots. A few magnified fractions of the images illustrate the symmetries in the field plots. The scalarized response function $f(z)$ is chosen as a linear function of the scattered field (Scheme 2.1 in Figure~\ref{fig:detailed_schematic}) under dipole excitation with variable position. When the source is placed in the resonator such that it possesses an even symmetry, modes of odd symmetry along the same axis are suppressed (gray lines indicate odd symmetry, green indicates even in the $x$-direction, orange indicates even in the $y$-direction). In the legend of the central figure, the axis is set in bold when the source is placed in a high symmetry position relative to that axis and a normal font when not. }
    \label{fig:selectve_excitation}
\end{figure}

Let us now move on to a more complex resonator comprising 30 scatterers. The scatterers are placed in \qty{200}{nm} steps to form two opposing meta-mirrors with a distance of \qty{5200}{nm}. The material model for silicon is replaced in favour of a nondispersive $\varepsilon_r = \complexnum{17.77 + 0.2j}$. All other parameters are as given in the first example. 
The resonator supports a rich spectrum of modes as shown in Figure~\ref{fig:selectve_excitation}. The central scatter plot contains the positions of the found resonances in the complex frequency plane for different excitations (indicated by different markers). The corresponding electric field distributions are illustrated above and below the scatter plot for two sets of resonances, which are grouped by the number of nodes along the horizontal $x$ direction (this grouping is also displayed in the scatter plot by the thick strokes). Even symmetries in the modal fields are indicated by a colored symmetry axis (green for $x$-, orange for $y$-symmetry), while a gray axis indicates odd symmetry. The three magnified regions shown on the right of Figure~\ref{fig:selectve_excitation} serve to visualize the described symmetries.

Robustly recovering the rich resonance spectrum involving a large number of modes is enabled by the adaptive refinement strategy described in Algorithm~\ref{algo:refinement}. It can be further helped by deliberately suppressing unwanted modes, which is possible by an appropriate choice of the scalarized response function $f(z)$. Here, $f(z)$ is chosen as one coefficient of the scattered field $p_\mathrm{local}$, under illumination from an electric dipole inside the cavity. The symmetry of the source inside the resonator is selectively broken by offsetting it from the center.
All modes couple to the source with broken symmetry in $x$ and $y$ (i.e., the grey $\times$ is present for all resonances). Excitations symmetric along one or more axes do not couple to modes antisymmetric along these axes. The orthogonality between excitation and mode becomes apparent by the absence of the respective pole. The divergent entries of the $\mathbf{T}_\mathrm{local}$-matrix are canceled in $\mathbf{T}_\mathrm{local}\mathbf{a}_\mathrm{local}$, while all entries of $\mathbf{a}_\mathrm{local}$ are non-zero. This can be understood as the extreme case of a vanishing residue due to a mismatch between excitation and mode. During the inverse design presented in the last section, we will analogously use the residue as a similarity measure between source and mode, to pre-select relevant resonances.



\subsection{Inverse Design}\label{sec:inverse_design}
\begin{figure}[ht]
    \centering
    \includegraphics[width=\linewidth]{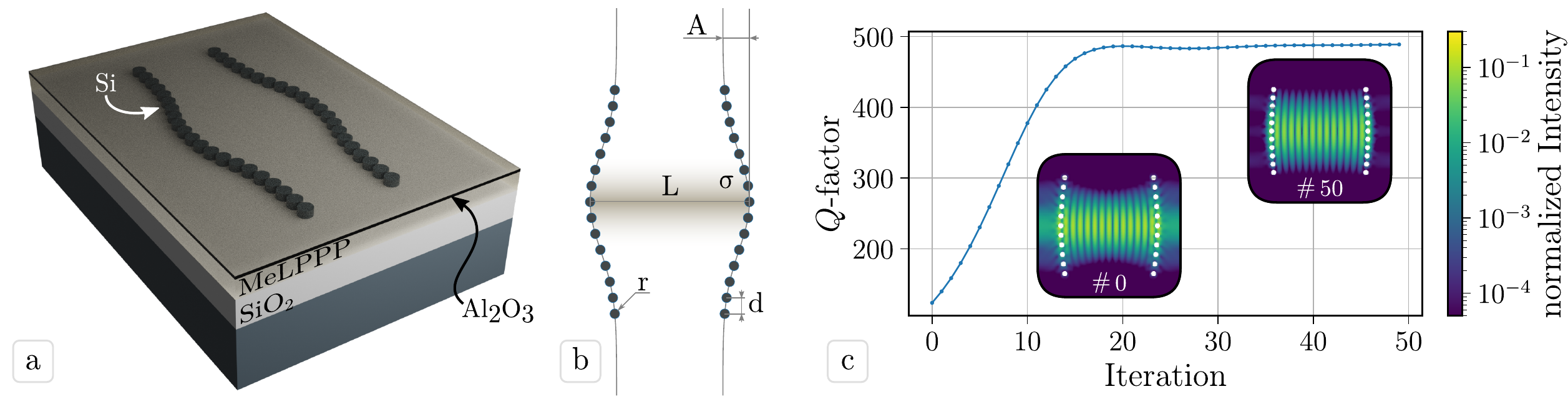}
    \caption{Inverse design of integrated polariton cavity formed by silicon cylinders for maximum $Q$-factor. The ladder polymer MeLPPP supports exciton-polaritons when placed in a resonator with sufficient $Q$. a) Layer stack for SOI chip integrated polariton cavity. A slab mode guided in the active MeLPPP layer is confined laterally by high contrast gratings formed from silicon posts. A thick layer of buried oxide isolates the resulting resonator from the silicon substrate. b) The posts' positions are parametrized by a Gaussian curve with variable amplitude $A$, standard deviation $\sigma$. Further free parameters are the center-to-center distance of the posts $d$ and the mirror distance at the cavity center $L$. The posts radius is fixed to $r=\qty{55}{nm}$. c) The evolution of the $Q$-factor with optimization iterations. Insets show the modal fields as an intensity distribution with a logarithmic colormap, allowing to clearly identify the radiative loss in the unoptimized geometry. After the optimization concludes barely any radiative loss remains visible.
    }
    \label{fig:optimization_sensitivities}
\end{figure}

As mentioned in Section~\ref{sec:pole_finding}, the AAA algorithm allows us to calculate the gradients of the resonance poles.
At the same time, gradients of the Mie coefficients and the TMF can be obtained via automatic differentiation. Thus, it is possible to efficiently differentiate the full solver chain with respect to, e.g., material parameters, geometry, and position of the scatterers. The availability of such gradients unlocks the efficient inverse design of modes of resonators made from multiple scatterers. Let us now demonstrate the inverse design of a photonic microcavity as employed for integrated polariton transistors \cite{tassanIntegratedUltrafastAlloptical2024}, with meta-mirrors consisting of silicon pillars. The top-down nanofabrication allows the accurate placement of the silicon pillars on demand. Therefore, the inverse design of the pillar arrangement opens the opportunity to suppress surplus optical losses of the photonic cavity resonance (i.e. increase the quality factor $Q = -\Re\{\tilde{z}\} / 2\Im\{\tilde{z}\}$ \cite{kristensenModesModeVolumes2014}), which is critical to enable exciton-polariton condensation \cite{tassanIntegratedUltrafastAlloptical2024}.

Making the layered structure shown in Figure~\ref{fig:optimization_sensitivities}a accessible to our method requires reducing it to constituents of high symmetry. Here, this is possible using an effective index approximation, which permits the description of the silicon posts as infinitely extended cylinders in a homogeneous background. As shown in Section~S6, the found modes and their relative positions agree with the experimental results for the right choice of effective dispersive permittivities. However, the effective index approximation has the limitation that out-of-plane radiation losses are not considered. Such a limitation could be circumvented phenomenologically, e.g., by adding a shell of a lossy material around each post. To ease reproducibility, we, however, keep the same permittivities as in the previous example, which approximately coincide with the calibrated effective refractive indices at $z=\qty{2.5}{eV}$.

To keep the complexity of this demonstration to a minimum, we will fix the posts' radius to $R=\qty{55}{nm}$, while the positions are parametrized to follow a Gaussian mirror shape inspired by \cite{tassanIntegratedUltrafastAlloptical2024} (see parameters indicated in Figure~\ref{fig:optimization_sensitivities}b). Indeed, the Gaussian parametrization is sufficient to reach $Q$-factors limited by non-radiative losses.  
The chosen objective function balances two sub-objectives by the weighting factor $\eta=\num{1e-2}$:
\[
F_\mathrm{obj} = \underbrace{\Im\{\tilde{z}_*\}^2}_\mathrm{primary} - \underbrace{\eta(\Re\{\tilde{z}_*\} - \omega_\mathrm{target})^4}_\mathrm{penalty}\, .
\]
The primary objective is to maximize the quality factor (or, equivalently, to minimize the distance between the pole and the real axis). A penalty term is added to keep the (real) resonance frequency approximately fixed, which ensures the resonance stays inside the search domain.
As $F_\mathrm{obj}$ solely depends on the position of the selected pole $\tilde{z}_*$, the sensitivities of all other poles do not contribute to the update step. 
The resonator geometry changes with every iteration, consequently modifying the resonance spectrum. The gradients generally differ between resonances. Thus, keeping track of the relevant resonance during the optimization is vital. A combination of three strategies accomplishes this. 
\begin{enumerate}
    \item As discussed in Section~\ref{sec:many_modes}, selected resonances are emphasized by an appropriate choice of $f(z)$.
    \item The resulting residues are combined with $\Im\{\tilde{\omega}_i\}$ to assign a relevance score to the poles in the considered spectral window (e.g., $\mathrm{Rel}(\tilde{z}_*) := \mathrm{Res}(\tilde{z}_*) / \Im\{\tilde{z}_*\}^2$).
    \item Lastly, to clearly identify the resonance of interest amongst the remaining most relevant resonances, we compare the modal profiles to a reference mode determined for the initial configuration of posts.
    For numerical efficiency, the comparison is performed by a small set of point evaluations. 
\end{enumerate}

The optimization is performed using the Adam stochastic gradient descent algorithm \cite{kingmaAdamMethodStochastic2017} with low momentum terms $\beta_1 = 0.8$ and $\beta_2 = 0.9$. 
Figure~\ref{fig:optimization_sensitivities}c shows the evolution of the quality factor $Q$ throughout the iterative optimization. The optimizer swiftly finds an optimized arrangement of posts, suppressing the in-plane radiation loss through the mirrors. Note how $Q$ does not increase monotonically due to the momentum terms of the Adam optimizer and the penalty term used to keep the resonance frequency's real part approximately fixed. The insets show how the radiative loss is clearly visible for the initial configuration, with little radiative loss after the optimizer has converged. Section~S7 of the SI discusses the evolution of design parameters throughout the optimization. The optimizer pushes the Gaussian tails outwards, approaching a parabolic mirror shape.
To further maximize $Q$ while reducing the resonator footprint, more degrees of freedom could be unlocked. Possible design freedoms include the individual placement of scatterers and their geometry. Core-shell cylinders or even freeform structures could replace the posts. Here, we refrain from such measures to guarantee compatibility with established fabrication techniques.

From additional simulations selectively suppressing loss channels, we conclude that the loss remaining for the final design is dominated by dissipation, predominantly in the silicon pillars. It should be noted that due to the effective index approximation out-of-plane radiation losses are not considered. To remedy this in the future, we suggest generalizing the proposed method to arbitrarily shaped objects  (i.e., cylinders of finite length) and their inclusion in stratified media, which are both compatible with the TMF \cite{beutelEfficientSimulationBiperiodic2021}. Careful treatment of the branch cuts emerging from a nonuniform background will be required. 

\section{Conclusion}
This contribution introduces an efficient method to calculate resonance frequencies and associated field profiles of composite resonators. We demonstrate the use of automatic differentiation, with appropriate derivatives for the AAA algorithm, to inversely design a composite resonator by maximizing its $Q$-factor. Further, selective excitation and post-selection of relevant modes are used to track the resonance of interest throughout the optimization. 

Currently, the proposed method applies to high symmetry scatterers (core-shell spheres and infinite cylinders), for which the $T$-matrix is known semi-analytically. An efficient method to evaluate the $T$-matrix of such scatterers for a high number of complex samples is needed to incorporate arbitrarily shaped scatterers. A representation of the $T$-matrix of an isolated scatterer in terms of a few relevant modes could potentially serve that purpose. Furthermore, the method proposed here is limited to finite clusters of scatterers. Numerous applications based on (locally) periodic arrangements demand a generalization to infinite lattice sums \cite{beutelEfficientSimulationBiperiodic2021}. Moreover, this would allow the incorporation of stratified embeddings/substrates by expanding the field into a finite set of plane waves. The developments presented in this contribution will serve as a foundation for future efforts addressing these open questions.


\pagebreak
\textbf{Acknowledgements} \par 
J.D.F. and C.R. acknowledge financial support by the Helmholtz Association in the framework of the innovation platform “Solar TAP”. 
C.R. acknowledges support from the German Research Foundation within the Excellence Cluster 3D Matter Made to Order (EXC 2082/1 under project number 390761711) and by the Carl Zeiss Foundation. 
N.A. and C.R.~acknowledge support by the Federal Ministry of Education and Research (BMBF) within the project DAPHONA (16DKWN039).
T.J.S. acknowledges funding from the Alexander von Humboldt Foundation.
F.Bi., F.Be., and S.B.~acknowledge funding
by the Deutsche Forschungsgemeinschaft (DFG, German Research Foundation) 
under Germany's Excellence Strategy - The Berlin Mathematics Research
Center MATH+ (EXC-2046/1, project ID: 390685689) and 
by the German Federal Ministry of Education and Research
(BMBF Forschungscampus MODAL, project 05M20ZBM). 
P.T., D.U., T.S., and R.F.M acknowledge funding from EU H2020 EIC Pathfinder Open project “PoLLoC” (Grant Agreement No. 899141) and EU H2020 MSCA-ITN project “AppQInfo” (Grant Agreement No. 956071).
J.D.F., F.Be., F.Bi., and S.B. acknowledge fruitful discussions with Martin Hammerschmidt and Lin Zschiedrich.
\medskip

%

\bibliography{article}

\end{document}


\mathtoolsset{centercolon}

\pagestyle{fancy}

\title{Supplementary Information: A framework to compute resonances arising from multiple scattering}
\maketitle


\author{Jan David Fischbach*}
\author{Fridtjof Betz}
\author{Nigar Asadova}
\author{Pietro Tassan}
\author{Darius Urbonas}
\author{Thilo Stöferle}
\author{Rainer F. Mahrt}
\author{Sven Burger}
\author{Carsten Rockstuhl}
\author{Felix Binkowski}
\author{Thomas Jebb Sturges}


\dedication{}










\begin{abstract}
\end{abstract}

\renewcommand{\thepage}{S\arabic{page}}
\renewcommand{\thesection}{S\arabic{section}}
\renewcommand{\thetable}{S\arabic{table}}
\renewcommand{\thefigure}{S\arabic{figure}}
\renewcommand{\theequation}{S\arabic{equation}}

\setcounter{figure}{0}
\setcounter{section}{0}

\section{Meromorphic T-matrix Formalism}
\label{sec:holomorphic}
\subsection{Mie Coefficients}\label{sec:meromorphic_mie}
The so-called (generalized) Mie coefficients are expressed as fractions of sums of (spherical) Bessel and Hankel functions. The argument to these is the complex-valued size parameter $x_\pm = k_\pm r$, where $r$ is the radius of the scatterer and $k_\pm$ is the wavenumber in the scatterer's medium, dependent on material parameters and frequency \cite{beutelHolisticFrameworkElectromagnetic2024}.
For lossless isotropic materials characterized by permittivity and permeability, the argument to the Bessel and Hankel functions has the same phase as the complex frequency ($\angle x_\pm = \angle z$).
The Bessel and Hankel functions of the first and second type are holomorphic on $\mathbb{C}$ cut along the negative real axis $(-\infty, 0]$ \cite[Section 10.2(ii)]{olverNISTHandbookMathematical2010}. Consequently, the Mie coefficients are meromorphic on the discussed domain of holomorphicity of the Bessel and Hankel functions.
For lossy media, the branch cut will be tilted according to the corresponding loss tangent.

\subsection{Translation Coefficients}\label{sec:holomorphic_C}
The coefficients in $\mathbf{C}^{(3)}$ translating the scattered field from one scatterer into the basis of multipoles incident on another scatterer are at the heart of Equation~3 of the main text. 
The expressions for the elements of $\mathbf{C}^{(3)}$ are given in
[Appendix C]\cite{beutelTreamsTmatrixbasedScattering2024}. For spherical waves, these are weighted sums of spherical Hankel functions ([Equations C.1, C.3]\cite{beutelTreamsTmatrixbasedScattering2024}). The spherical Hankel functions can be obtained from the Hankel functions by:
\[
h_n^{(1,2)}(x)=\sqrt{\frac{\pi}{2x}}H_{n+ \frac{1}{2}}^{(1,2)}(x)\, .
\]
As such, they are holomorphic on $\mathbb{C}$ cut along the negative real axis $(-\infty, 0]$, provided the branch cut of $\sqrt{\cdot}$ is chosen accordingly.

For cylindrical waves, the coefficients are given by [Equation~C.5]\cite{beutelTreamsTmatrixbasedScattering2024}:
\[C_{m, m^\prime}^{(3)}(k_\pm, \mathbf{r}-\mathbf{r}^\prime) = 
H_{m-m^{\prime}}^{(1)}\left(\sqrt{k_\pm^{2}-k_{z}^{2}}\rho_{r-r^{\prime}}\right)\mathrm{e}^{\mathrm{i}(m-m^{\prime})\varphi_{r-r^{\prime}}+\mathrm{i}k_{z}(z-z^{\prime})}\, ,\]
where $\rho_{r-r^{\prime}}$, $\varphi_{r-r^{\prime}}$ and $z-z^{\prime}$ relate original and translated basis sets. In this contribution $k_z=0$, as we are interested in stationary modes. As a consequence, the argument of $H_{m-m^{\prime}}^{(1)}$ reduces to $k_\pm\rho_{r-r^{\prime}}$ so that the domain of holomorphicity is the same as discussed for the Mie coefficients (Section \ref{sec:holomorphic}).

\section{AAA Autograd Derivatives}
\label{sec:derivatives_jax}
Here, we will briefly elaborate on how the derivatives introduced in  \cite{betzEfficientRationalApproximation2024} are implemented as JAX-compatible Jacobian Vector products (JVPs) in \emph{diffaaable}.

Given the tangents $\frac{\partial f_\mathrm{k}}{\partial \varrho}$, we will use the chain rule on $r(w_j, f_j, z)$ along its weights $w_j$ and values $f_j$ (the nodes $z_j$ are treated as independent of $\varrho$) \cite[Equation~4]{betzEfficientRationalApproximation2024}:
\[
\frac{\partial f_\mathrm{k}}{\partial \varrho} \approx \frac{\partial r_\mathrm{k}}{\partial \varrho}= \sum_{j=1}^m\frac{\partial r_\mathrm{k}}{\partial f_j}\frac{\partial f_j}{\partial \varrho}+\sum_{j=1}^m\frac{\partial r_\mathrm{k}}{\partial w_j}\frac{\partial w_j}{\partial \varrho}\, .
\]

To solve this system of equations for $\frac{\partial w_j}{\partial \varrho}$, we express it in matrix form:

\[\mathbf{A}\mathbf{w}^\prime = \mathbf{b}\, ,\]

where $\mathbf{w}^\prime$ is the column vector containing $\frac{\partial w_j}{\partial \varrho}$ and $\mathbf{b}$ and $\mathbf{A}$ are defined element wise: 

\[
\begin{aligned}
    b_\mathrm{k} &= \frac{\partial f_\mathrm{k}}{\partial \varrho} - \sum_{j=1}^m\frac{\partial r_\mathrm{k}}{\partial f_j}\frac{\partial f_j}{\partial \varrho}\\
    A_{kj} &= \frac{\partial r_\mathrm{k}}{\partial w_j}\, .
\end{aligned}
\]

These are augmented with Equation~5 of \cite{betzEfficientRationalApproximation2024}, which removes the ambiguity in $\mathbf{w}^\prime$ associated with a shared phase of all weights.

The expressions for the derivatives of $r_\mathrm{k}$ are found in the definition of $r(z)$ (main paper Equation~4): \[ \frac{\partial r_\mathrm{k}}{\partial f_j}= \frac{1}{d(z_\mathrm{k})} \frac{w_j}{z_\mathrm{k}-z_j}\] and
\[
\frac{\partial r_\mathrm{k}}{\partial w_j}= \frac{1}{d(z_\mathrm{k})} \frac{f_j-r_\mathrm{k}}{z_\mathrm{k}-z_j} \approx \frac{1}{d(z_\mathrm{k})} \frac{f_j-f_\mathrm{k}}{z_\mathrm{k}-z_j}\, .
\]


\section{Comparison to Arnoldi Iteration}
\label{sec:errorT}

\begin{figure}[ht]
    \centering
    \includegraphics[width=\linewidth]{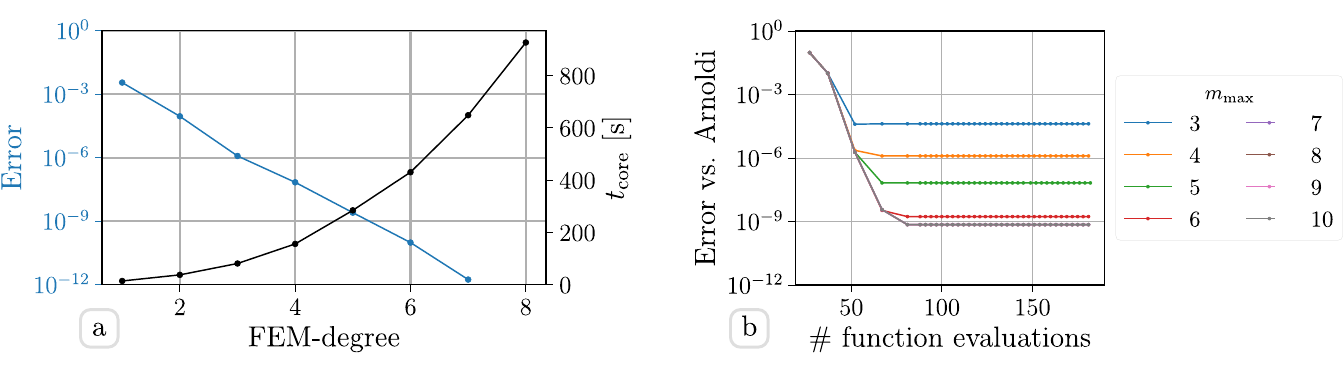}
    
    \caption{Comparison to the state-of-the-art reference solver. a) Convergence and computational cost of the Arnoldi iteration with FEM degree. The same resonator as in Section~3.1 of the main text is considered. The geometry is discretized with a fixed mesh of \qty{24}{nm} maximum side length and fixed PML boundary conditions are used, while the FEM-degree is swept. b) Recreation of Figure~3b of the main text with the Arnoldi results as reference. The converged resonance frequencies from a) are used as the reference to compute the maximum relative error amongst 13 compared frequencies (denoted as ``Error").}
    \label{fig:convergence_poles_vs_arnoldi}
\end{figure}

Here, we compare resonance frequencies and field distributions obtained by the proposed method to an established reference solver.
The reference solutions are found by linearizing the NEP via auxiliary fields. The resulting linear eigenvalue problem is then solved via Arnoldi iteration. The convergence plot analogous to Figure~3b of the main text is shown in Figure~\ref{fig:convergence_poles_vs_arnoldi}b. Note the limiting error of $\sim 10^{-9}$, despite the smaller error $\sim 10^{-11}$ between Arnoldi results for different FEM-degree. This is likely caused due to the inaccuracies in representing the geometry by finite elements for the given mesh size (\qty{24}{nm} maximum side length) and curvilinear-degree (5).

\begin{figure}[ht]
    \centering
    \includegraphics[width=\linewidth]{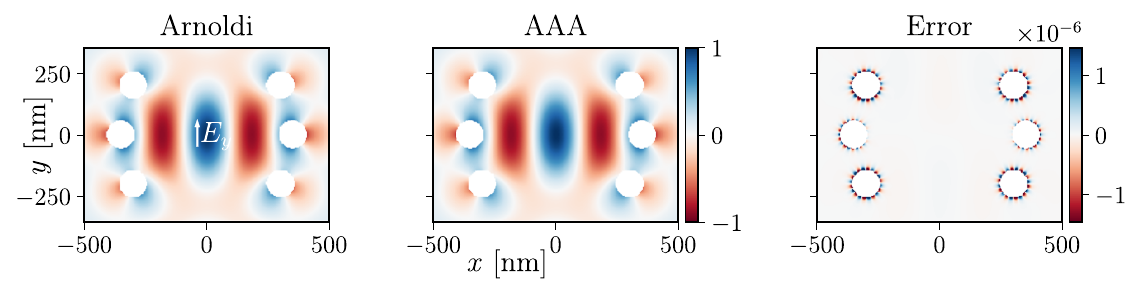}
    \caption{Comparison of the modal profile w.r.t.~a reference solution obtained by Arnoldi iteration. Both modes have been scaled such that $E_y(x=0, y=0) = 1$. The difference $E_{y,\mathrm{Arnoldi}}-E_{y,\mathrm{AAA}}$ is concentrated around the cylinders and has $m_\mathrm{max}+2=10$ angular oscillations.}
    \label{fig:field_convergence}
\end{figure}

Figure~\ref{fig:field_convergence} compares the field distribution found using the proposed method and the Arnoldi iteration. As discussed above, the error is concentrated around the scatterers and reduces with $m_{max}$.

Increasing $m_{max}$ further leads to the reappearance of errors, attributed to the numerical discretization of the Bessel and Hankel functions, similar to Section~3.1 of the main text. It is noted that evaluating the modal fields by a single scattering simulation is less affected by this phenomenon than the weighted superposition. A possible explanation for this behavior is the argument dependence of the numerical stability of the used implementation of the Hankel functions \cite{amosAlgorithm644Portable1986}. 

\section{Dispersive Material Model Silicon Posts}\label{sec:silicon}
In the first example we treat dispersive materials to demonstrate the compatibility of our method. The dispersion is modelled by a Lorentz oscillator model of the form:
\[
\varepsilon_r(z) = 1 + \sum_l \frac{is_l}{z-z_{p, l}} + \frac{is_l^*}{z+z_{p, l}^*}
\].

The corresponding parameters for the used silicon material model with two pole pairs (angular frequencies) are $(z_{p,1}, s_1) = (\num{6.29982176e15} - \num{8.75108242e14}i, \num{1.67145513e15} + \num{2.93759023e16}i$) and $(z_{p,2}, s_2) = (\num{5.12113206e15} - \num{2.18905716e14}i,\num{4.16863154e15} + \num{4.24772028e15}i)$.

\section{Scattering Problem: Comparison to FEM}
\begin{figure}
    \centering
    \includegraphics[width=0.25\linewidth]{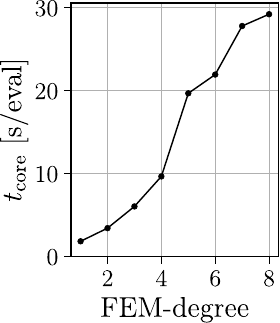}
    \caption{The computational cost to solve a single finite element method (FEM solever JCMsuite) scattering problem. The same resonator as discussed in Figure~3 of the main text is considered. It is discretized with the same mesh as used for Figure~\ref{fig:convergence_poles_vs_arnoldi}}
    \label{fig:scat_fem}
\end{figure}

As discussed in the introduction the second approach to finding resonances - namely finding singularities of the scattering problem - enables the use of any solver suitable to evaluate the analytic continuation of the scattering problem at complex frequencies. For suitable structures this evaluation is possible via the TMF, as discussed in the main text. In Figure~\ref{fig:scat_fem} we report on the computational cost of evaluating the scattering problem using FEM for different FEM-degree, which maps to an accuracy when determining the resonance frequencies as seen in Figure~\ref{fig:convergence_poles_vs_arnoldi}.

\section{Polariton Cavities: Match to Experiment}\label{sec:calibration}
\begin{figure}[ht]
    \centering
    \includegraphics[width=0.5\linewidth]{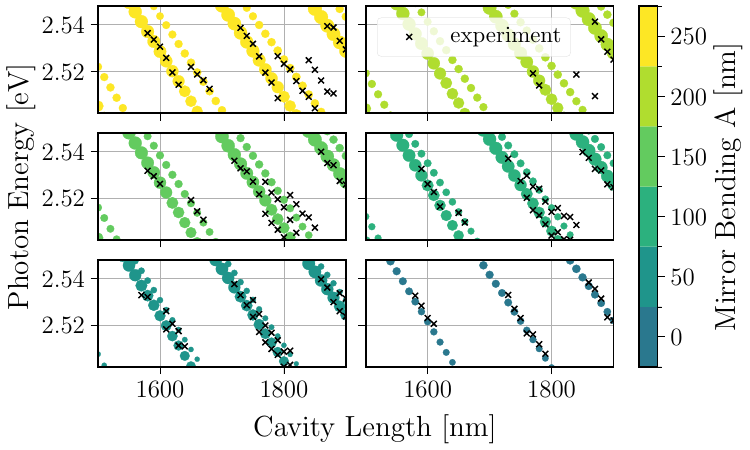}\hspace{1cm}
    \includegraphics[width=0.3\linewidth]{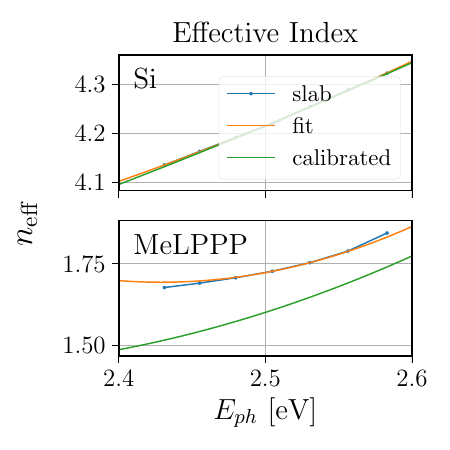}
    \caption{The effective index approximation used to represent the polariton cavities by infinitely extended cylinders is compared to experimental results \cite{tassanIntegratedUltrafastAlloptical2024}. Left: The first few transversal modes per longitudinal mode are plotted for different cavity length and mirror bending. Good agreement to the experimental data indicated in black is reached by adjusting the coefficient of a second order polynomial fit to the slab modes of the silicon and MeLPPP stacks. Right: The corresponding effective indices of a TE polarized slab mode (blue), its polynomial fit (orange), and the adjusted fit to yield experimental agreement. }
    \label{fig:experiment}
\end{figure}

As mentioned in Section~3.4, we use the effective index approximation to reduce the 3D layer stack shown in Figure~5a to an effective 2D arrangement of infinite cylinders. To demonstrate the ability to accurately represent photonic properties of the actual 3D geometry despite the approximation, we show how the found resonances correspond the experimental data for appropriately chosen indices (and their dispersion). The considered experimental data \cite{tassanIntegratedUltrafastAlloptical2024} was obtained extracting the (real-valued) resonance frequencies from the emission spectra of optically pumped exciton-polariton (polariton) cavities in the regime where polariton condensation occurs. As such the experimentally recorded resonance frequencies are affected by the blue-shift characteristic to polariton condensation. The blue-shift can be accounted for to the zeroth-order by calibrating the $n_\mathrm{eff}$ in the MeLPPP region and its dispersion to faithfully reproduce the experimental observations.
Initial estimates were taken as the effective indices of the slab modes, those were fitted by their second order Taylor expansion. The coefficients were subsequently adjusted to yield a match to experiment as seen in Figure~\ref{fig:experiment}.

\section{Optimization: Parameter Space Evolution}\label{sec:evolution}

\begin{figure}[ht]
    \centering
    \includegraphics[width=0.7\linewidth]{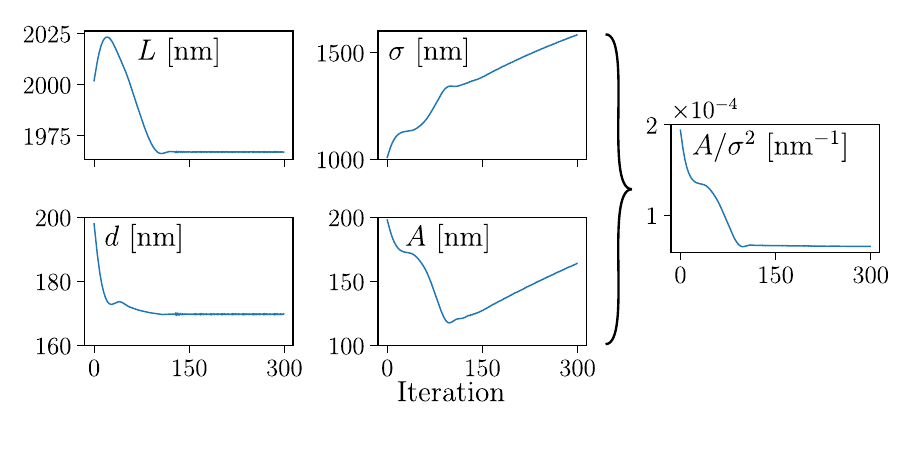}
    \caption{Evolution of the Gaussian resonator parameterization (shown in Figure~5b) during optimization. In contrast to Figure~5c, the optimization is extended to 300 iterations, which does not yield significant further improvement in $Q$ (not shown).}
    \label{fig:parameter_evolution}
\end{figure}
In this section, we study the variation of the parameters characterizing the cavity during the optimization. The results corresponding to the optimization shown in Figure~5c are shown in Figure~\ref{fig:parameter_evolution}.
During the optimization, the cavity length $L$ and spacing between cylinders $d$ quickly converge (Figure~\ref{fig:parameter_evolution} left column). For a Fabry-Perot resonator with uniform mirrors, the cavity length determines its resonance frequency, while for a fixed cavity length, the reflectivity governs the quality factor of the cavity. We conclude that $L$ converges to fulfil the $\Re\{\tilde{z}_*\} - \omega_\mathrm{target}$ sub-objective, while $d$ is optimized to maximize the mirror reflectivity (in agreement with \cite{tassanIntegratedUltrafastAlloptical2024}).
The Gaussian amplitude $A$ and width (quantified as its standard deviation $\sigma$) do not converge in the course of the optimization (Figure~\ref{fig:parameter_evolution} center column). Interestingly, the influence on the arrangement of posts is minimal in the center, increasing towards the tails of the Gaussian, which progressively get pushed away from the mirror center. As the number of cylinders is fixed, the influence of the Gaussian tails progressively vanishes, leaving us with the central lobe only. Let us describe its shape by the second-order Taylor expansion of the Gaussian parameterization:

\[
\begin{aligned}
g(y) &= L/2 - A + A e^{-y^2/2\sigma^2}\\
\mathrm{with}\\
g^\prime(y) &= -\frac{A y e^{-y^2/2\sigma^2}}{\sigma^2} = -\frac{y f(y)}{\sigma^2}\\
g^{\prime\prime}(y) &= \frac{f(y)\left(y^2 - \sigma^2\right)}{\sigma^4} \, .
\end{aligned}
\]

The Taylor expansion around $y=0$ thus reads as:

\[
\begin{aligned}
t(y) &= L/2 - A/2\sigma^2 y^2\, .
\end{aligned}
\]

As shown in the right column of Figure~\ref{fig:parameter_evolution}, the curvature $A/\sigma^2$ has a limiting value during the optimization contrary to $A$ and $\sigma$. The presented inverse design thus tends towards parabolic mirrors as the limiting case of the adopted Gaussian shape \cite{tassanIntegratedUltrafastAlloptical2024}.

\bibliography{article}